\documentclass[journal]{IEEEtran}
\ifCLASSINFOpdf
  \usepackage[pdftex]{graphicx}
  \graphicspath{{../pdf/}{../jpeg/}}
  \DeclareGraphicsExtensions{.pdf,.jpeg,.png}
\else
\fi
\usepackage{amsmath}

\usepackage{subfig}
\usepackage{amssymb}

\usepackage{tikz}
\usetikzlibrary{shapes,arrows}

\begin{document}
%
\title{Intermediate frequency Upgrade design features of NASA D3R Weather Radar System}
%
%
%

\author{Mohit Kumar,~\IEEEmembership{Student Member,~IEEE,}, Shashank Joshil,~\IEEEmembership{Student Member,~IEEE,}, Manuel Vega,~\IEEEmembership{Member,~IEEE,}, Robert Beauchamp,~\IEEEmembership{Member,~IEEE,}
        and V Chandrasekar,~\IEEEmembership{Fellow,~IEEE}
}

\maketitle

\begin{abstract}
The NASA dual-frequency, dual-polarization, Doppler radar (D3R) is an
important ground validation tool for the global precipitation measurement (GPM) mission’s dual-frequency precipitation radar (DPR). The D3R has undergone extensive field trials starting in 2011 and continues to provide observations that enhance our scientific knowledge. To further enhance its capabilities, the Intermediate frequency (IF) electronics, digital receiver and waveform generation subsystems have been upgraded. Due to the new, more flexible architecture, this upgrade enables more research frontiers to be
explored with better performance. One of the primary motivations for this upgrade is to enable enhanced radar sensitivity and increase range resolution to 30 meters.  In this work, the D3R system’s upgrade will be discussed with a focus on the key upgrade's design features to obtain better sensitivity and a flexible waveform capability. 
\end{abstract}


%
\IEEEpeerreviewmaketitle

\section{Introduction}
%
%
%
%
\IEEEPARstart{T}{he} NASA dual frequency, dual polarization doppler radar (D3R) was developed as a joint collaboration between NASA, Colorado state University and Remote Sensing Solutions. It allows for beam aligned, synchronized operation between Ku band (13.91GHz) and Ka band (13.51GHz) for observation of an meteorological event. D3R forms an important ground validation tool for dual precipitation radar (DPR) for the GPM mission and has undergone extensive field trials earlier (\cite{First5yrs} and \cite{FiveYrsOp}). The Ku band can sample volume upto 40 Kms with peak power of 160 Watts from solid-state transmitter and similarly Ka band has peak power of 40 Watts, providing high sensitivity to light rain and snow. D3R also uses a multi-pulse scheme to mitigate the effects of blind range caused by longer pulses and subsequently using pulse compression to enhance the range resolution and sensitivity \cite{Vega2014}. \par

The use of solid-state transmitters together with receive systems with pulse compression waveforms enable flexible architecture wherein high-end signal processing techniques like digital beam-forming and Space Time Adaptive processing (STAP) algorithms can be utilized. However, a major source of error for volume target or weather sensing for these pulse compression systems, can be the side-lobe contamination in adjacent range gates. Mismatched filtering techniques have been known to reduce peak auto-correlation sidelobe levels \cite{Bharadwaj2012} and has been used in this work. The current system lacked programmability and reconfiguration options.\par

The hardware upgrade focused on making the system more robust and flexible to adaptations which the modern weather systems need \cite{Kumar8517944}, \cite{8128188}. The pulse compression filter supports 480 taps (multiply-accumulate operations, MACs) to accommodate the pulse compression sidelobe energy to higher number of taps due to the mismatched filtering approach that we adopted. These large number of multipliers for both in-phase and quadrature components for the horizontal and vertical polarization were made possible due to the high number of MAC units in Virtex 6 FPGAs enabled through floor planning and advanced synthesis constraints. The hardware upgrade will open up new research frontiers in weather science and engineering at Ku and Ka band for developing accurate models and algorithms for the GPM core satellite's DPR. \par

Also with the hardware upgrade, the resolution can be increased to 30m from the current configuration of 150m. Additionally, with a coupled waveform generator, digital down-converter and pulse compressor design, pulse by pulse change of waveform and filters, can be achieved synchronously. \par

The timing portion of D3R is setup in such a way that the acquired samples have noise injection, for receive calibration, followed by short and medium pulse injection into receive (for transmit calibration). Both noise source and transmit signal injection aids in online monitoring of transmit and receive portions of the radar. The short pulse is not modulated and it only goes through a delay filter but the medium pulse is modulated with Chirp signal and it goes through pulse compression filters (FIR based). With a good amount of power of the medium pulse through the calibration loop, there is likelihood of receive samples getting submerged with the sidelobes of transmit pulse (through the calibration loop), if the pulse compression filters are long. Mismatched filtering technique often leads to longer pulse compression filters to suppress the Integrated sidelobe levels (ISL). However, this can severely limit the weak echoes near to the transmit calibration signal. Hence the sidelobes need to be sufficiently low power in terms of peak values and ISL.  This is also true for the case of strong returns from weather as well, wherein the sidelobe power, if not properly contained, can submerge weak echoes in range cells even far off, due to large lengths of these filters.\par

Arbitary Waveform Generator and digital receive sub-systems were upgraded along with the associated IF electronics. Since there was a need to change the 1st IF, hence the clock sections and the IF up-converters and down-converters were also modified. Both the IF waveform generator and the digital receive sections have been packaged into the XMC module based on virtex 6 FPGAs. This module includes three 200MHz, 16bit A/Ds, two 800 MHz D/As and four banks of memory. The ADCs and DACs have a high spurious free dynamic range (SFDR), so as to able to observe large storms to light cloud and snow. With high dydnamic range, the sensitivity of the system can be improved, if the receive gain is tailored to toggle the LSB of the A/D converter. This will enable the noise to be contained in the lowest bits. However, it requires fine tuning of the receiver gain. Usually, if the SFDR is not good, additional spurious signals can also get detected along with the weather echoes. These spurious signals might be due to the inter-modulation products of the mixer, which comes in band due to aliasing or improper filtering or due to non-linearity of the A/D itself. In this work, we present an analysis as to how to contain these inter-modulation products and demonstrate the filtering requirements for such a high dynamic range receiver.

\section{Hardware Upgrade features}
The XMC module with DAC interface can operate upto 80 MHz to maintain reasonable sampling rates with which digital in-phase and quadrature samples can be interfaced from FPGA to DAC device and also to have a feasible filter design. We redesigned the IF stages and selected an analog bandwidth with simulations of the mixer products and their suppression in order to meet the dynamic range requirement. Main challenging part here was the removal of the local oscillator (LO) feed through (which is quite high power because of the double balanced mixer and its need of high LO power). This is usually a trade-off with other inter-modulation products. Hence we came up with the analog bandwidth number which gives adequate rejection of LO feed through, other mixer products and harmonics.\par

\begin{figure}[!t]
	\centering
	\includegraphics[width=3.5in]{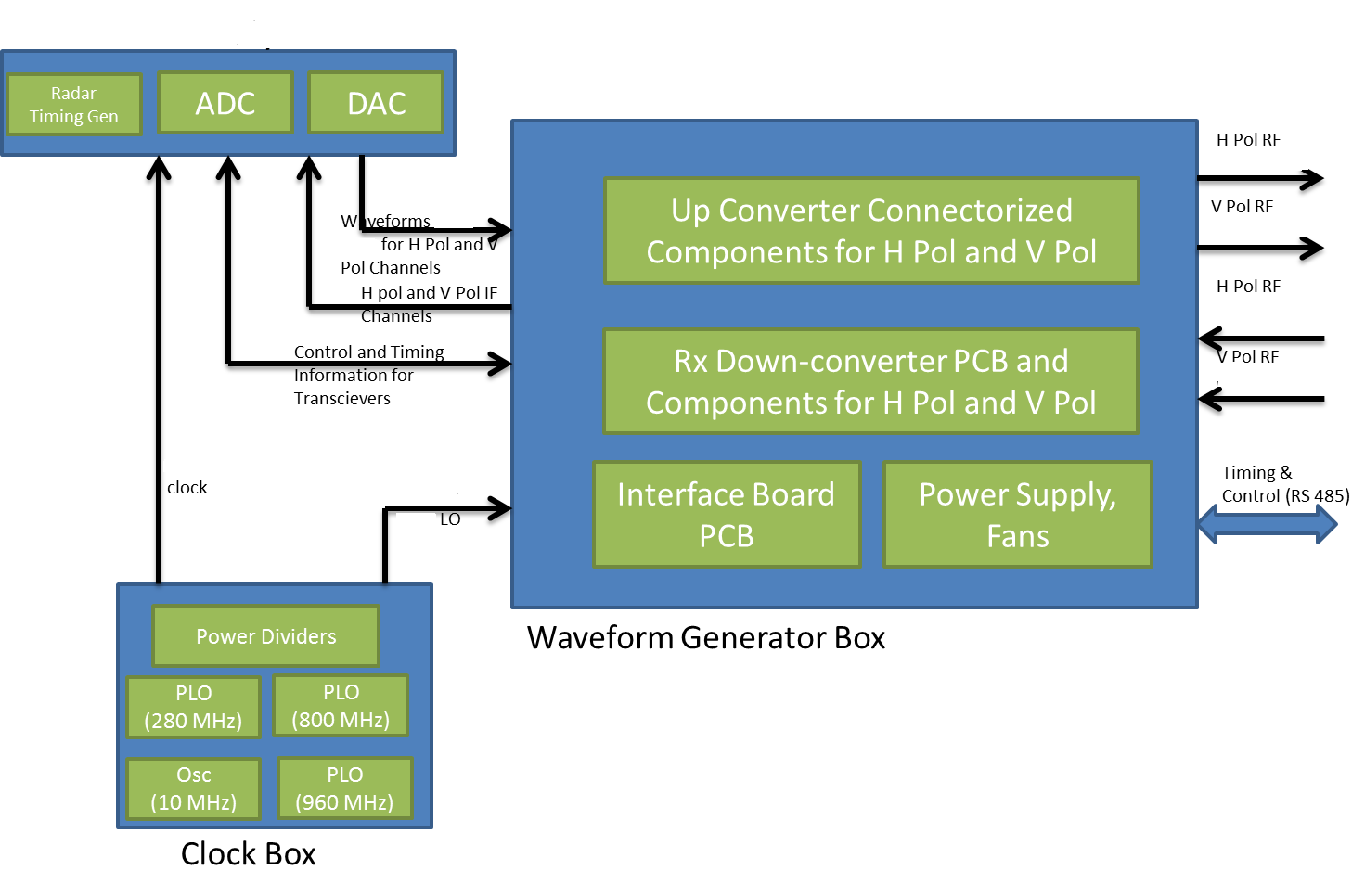}
	\caption{Overall segregation of functions into different sub-systems.}
	\label{fig_sim1}
\end{figure}

The initial design activity started with exploring multiplier approach , which could enable us to retain the existing IF hardware. The idea was to use a frequency doubler and its input would be half band frequencies from the XMC module. But this idea was short lived as we observed pulse droop and spurious at the output of the multiplier. Later we went for mixer approach and for the design of the IF stages with mixer, we undertook the simulation of the spurious of the mixer and suppression due to filters. By comparing different mixer parts from companies along with the simulation of spurious and filter characteristics, we could finalize the mixer part. The selected mixer had approximately 6 dB conversion loss, 30dB LO to IF isolation, 60dB suppression of 2nd and 3rd order inter-modulation products and LO drive level of 17dBm. To adjust to the changes in IF value, the IF systems was redesigned preserving the form factor for the existing sub-systems and modifying them with different components internally. Also we wanted minimal changes in cabling and power. The overall design functions are in Fig. \ref{fig_sim1}.   

\subsection{IF Up-Converter}
With the simulations carried out for the mixer intermodulation products and the IF gain required, the design evolved for the IF up-converter portion for the Ku and Ka band. The overall design philosophy of segregating the LO portions to clock box and IF up-convert stages to Waveform box (these names come from existing design), was retained. The LO power requirements of +17dBm were satisfied partially at the clock box level and at the waveform generator box level. The overall design met the drive requirement of the solid-state power amplifier inputs and maintained a good spectral purity of signals (through adequate filtering). The filter specifications were more tight for the up-convert stages than the receive, because the power flowing through each of the up-convert stages are high and most of the components in that path operate near to the 3dB compression levels (in the IF up-converter), \cite{kumar2019receive}.
\begin{figure}[!t]
	\centering
	\includegraphics[width=2.5in]{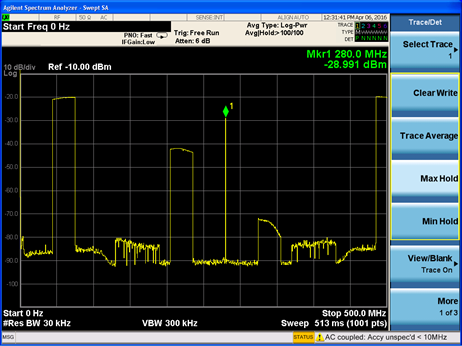}
	\caption{Inter-modulation products of the mixer device}
	\label{fig_sim4}
\end{figure}

\subsection{IF Receiver Design Features}
The IF receiver path components starts with band-pass filter, followed by an attenuator and the blocking capacitor.
The amplifier used after this stage is a flat gain and low noise for the band of interest. Next a low-pass filter is utilized to suppress the harmonics from the amplifier. The LO path components have attenuators and amplifiers, which would finally feed the LO port of the mixer. The Amplifier has an in built stabilization network. The traces on the printed circuit board (PCB) have line widths designed to make them controlled impedance. The power consumption of the amplifier is close to 0.35 W. The PCB thickness is 60 mils and plating thickness of 0.8 mils, hence for a 13 mil via with a thermal resistance of 200 deg C/Watt would give a the temp raise approx 70 deg C. For multiple via holes, the thermal resistance decreases proportionately and the temp raise would much less. The various inter-modulation products of the mixer are shown in Fig. \ref{fig_sim4}. The high value spurious products shown get suppressed by filters before A/D converter device.  \par
There is another amplifier that we used in the recieve path whose 1dB Compression point ensures that the A/D is not damaged in the long run with higher power echoes reflected from near ranges or ground clutter. \par

All the components in the receive path have high third order intercept point (OIP3) such that the resultant OIP3 is quite high. This aided us in achieving a high SFDR (Spurious free dunamic range). The components with high OIP3 consume more power and get more hot, thus there is a trade-off between power consumption and OIP3.

\begin{figure}[!t]
	\centering
	\includegraphics[width=2.5in]{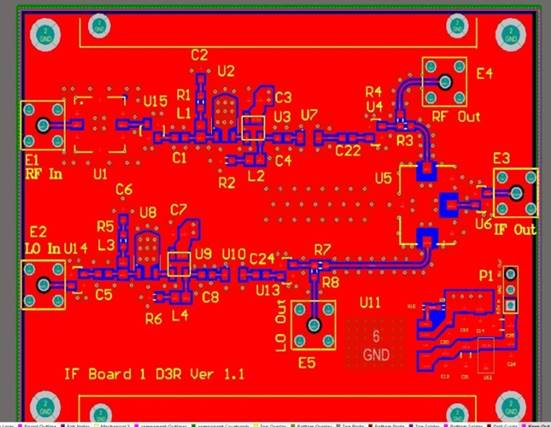}
	\caption{The PCB with mixer device layout}
	\label{fig_sim5}
\end{figure}

\begin{figure}[!t]
	\centering
	\includegraphics[width=2.5in]{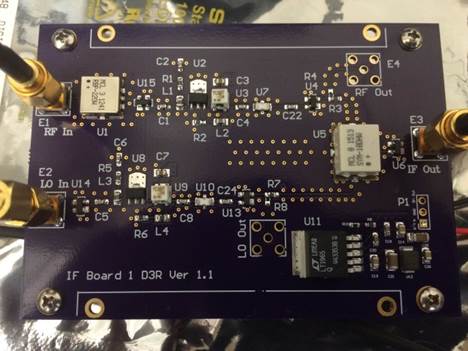}
	\caption{The final fabricated and assembled PCB for mixer device.}
	\label{fig_sim6}
\end{figure}

\subsubsection{Board layout}
\begin{figure}[!t]
	\centering
	\includegraphics[width=2.5in]{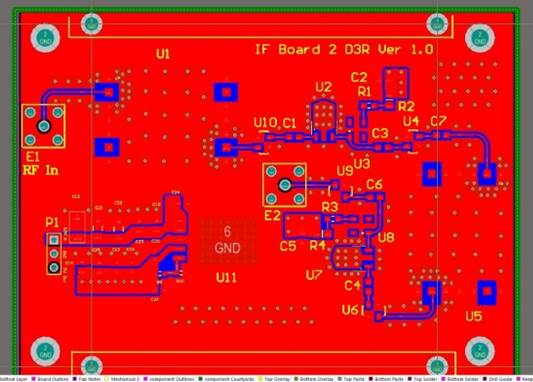}
	\caption{The amplifier PCB board layout}
	\label{fig_sim7}
\end{figure}

\begin{figure}[!t]
	\centering
	\includegraphics[width=2.5in]{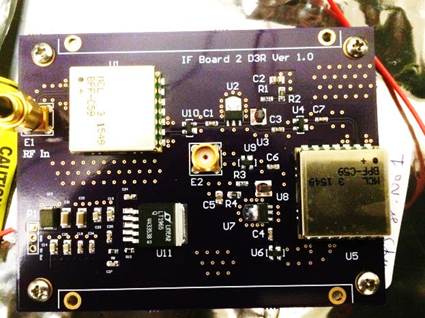}
	\caption{The realized and assembled amplifier device board.}
	\label{fig_sim8}
\end{figure}

The LO and RF paths were made symmetric and the line lengths were kept small compared to wavelength so as to minimize emissions. The trace widths were calculated keeping in mind the layer stackup and the dielectric constant of the substrate. All of the PCBs had four layer stackup. The whole of top and bottom layers are kept for ground having islands for power. We used via stitching for power and ground planes to minimize the loop currents. The board layout and the fabricated boards are shown in Fig. \ref{fig_sim5}, \ref{fig_sim6}, \ref{fig_sim7} and \ref{fig_sim8} respectively.

\subsubsection{Enclosure Shielding for IF Receiver boards}
\begin{figure}[!t]
	\centering
	\includegraphics[width=3.5in]{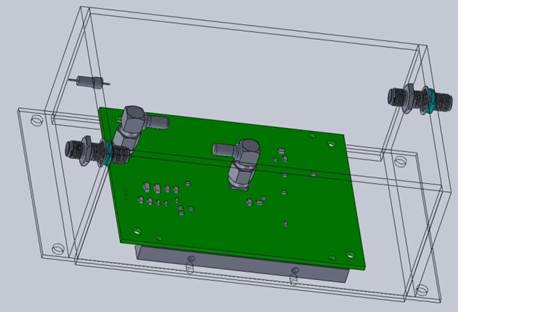}
	\caption{Simulated shielding enclosure.}
	\label{fig_sim9}
\end{figure}

The IF shield enclosures are metal cases with 3 in width, 5 in length and 2 in height to reduce any EMI emanating from the designed boards as well as to protect them from radiated emissions from elsewhere. They were designed with mounting bars and bulkhead panel mount SMA connectors with internal cabling. Prior to manufacture, the whole design was accomplished in SolidWorks. The simulated model is shown in Fig. \ref{fig_sim9}.

\subsection{Interface Board}
\begin{figure}[!t]
	\centering
	\includegraphics[width=2.5in]{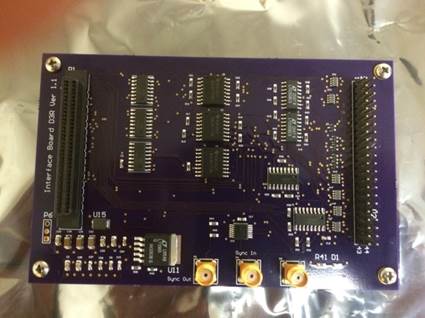}
	\caption{The PCB layout of the interface board}
	\label{fig_sim10}
\end{figure}

\begin{figure}[!t]
	\centering
	\includegraphics[width=2.5in]{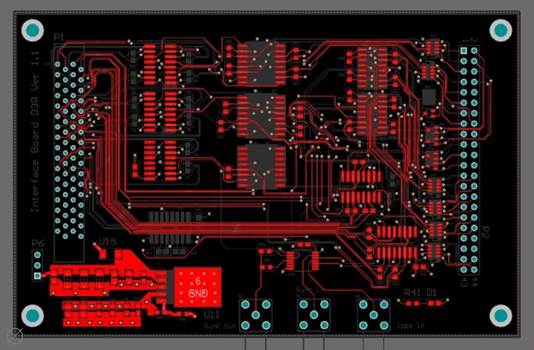}
	\caption{The fabricated and assembled PCB}
	\label{fig_sim11}
\end{figure}

It was used to convert signals from LVDS interface to RS485 interface. Transient suppressors were added to the RS485 lines which would traverse long distance. The sync in and sync out were on the coaxial interface. The sync in and sync out were provided so as to maintain synchronization between Ku and Ka systems. The board layout and manufactured board are shown in Fig. \ref{fig_sim10} and \ref{fig_sim11}. 

\subsection{Mechanical Aspects}
Three options were considered for design of Amplifier and Mixer board shield enclosures with different lengths (4,5 and 6in long). The 4 inch enclosure need the RF bulkheads to come out directly from the PCB. The 6 in enclosure was too big for our space budget. Best option was to use 5 in length box, where the bulkheads could be conveniently mounted on the walls.

\section{Conclusion}
The new IF hardware was developed and were integrated with the rest of the systems at the NASA Goddard Space Flight Centre and Wallops Flight facility, VA. The initial test results looked satisfactory and radar has been successfully deployed for weather monitoring of Winter Olympic games in PheongChang, South Korea in 2018 (\cite{8899120},\cite{Icepop1}) and in Wallops Flight facility later on. The software development to leverage the new facilities available in radar now are underway.




%





\ifCLASSOPTIONcaptionsoff
  \newpage
\fi
\bibliographystyle{IEEEtran}
\bibliography{references_arxiv1}

\begin{thebibliography}{1}
\providecommand{\url}[1]{#1}
\csname url@samestyle\endcsname
\providecommand{\newblock}{\relax}
\providecommand{\bibinfo}[2]{#2}
\providecommand{\BIBentrySTDinterwordspacing}{\spaceskip=0pt\relax}
\providecommand{\BIBentryALTinterwordstretchfactor}{4}
\providecommand{\BIBentryALTinterwordspacing}{\spaceskip=\fontdimen2\font plus
\BIBentryALTinterwordstretchfactor\fontdimen3\font minus
  \fontdimen4\font\relax}
\providecommand{\BIBforeignlanguage}[2]{{%
\expandafter\ifx\csname l@#1\endcsname\relax
\typeout{** WARNING: IEEEtran.bst: No hyphenation pattern has been}%
\typeout{** loaded for the language `#1'. Using the pattern for}%
\typeout{** the default language instead.}%
\else
\language=\csname l@#1\endcsname
\fi
#2}}
\providecommand{\BIBdecl}{\relax}
\BIBdecl

\bibitem{First5yrs}
V.~{Chandrasekar}, R.~M. {Beauchamp}, M.~{Vega}, H.~{Chen}, M.~{Kumar},
  S.~{Joshil}, M.~{Schwaller}, W.~{Petersen}, and D.~{Wolff}, ``Meteorological
  observations and system performance from the nasa d3r's first 5 years,'' in
  \emph{2017 IEEE International Geoscience and Remote Sensing Symposium
  (IGARSS)}, July 2017, pp. 2734--2736.

\bibitem{FiveYrsOp}
V.~{Chandrasekar}, H.~{Chen}, M.~{Vega}, R.~M. {Beauchamp}, M.~{Kumar},
  S.~{Joshil}, W.~{Petersen}, D.~{Wolff}, and M.~{Schwaller}, ``Observations
  and performance of the nasa dual-frequency dual-polarization doppler radar
  (d3r) from five years of operation,'' in \emph{2017 XXXIInd General Assembly
  and Scientific Symposium of the International Union of Radio Science (URSI
  GASS)}, Aug 2017, pp. 1--2.

\bibitem{Vega2014}
\BIBentryALTinterwordspacing
M.~A. Vega, V.~Chandrasekar, J.~Carswell, R.~M. Beauchamp, M.~R. Schwaller, and
  C.~Nguyen, ``Salient features of the dual-frequency, dual-polarized, doppler
  radar for remote sensing of precipitation,'' \emph{Radio Science}, vol.~49,
  no.~11, pp. 1087--1105, 2014. [Online]. Available:
  \url{https://agupubs.onlinelibrary.wiley.com/doi/abs/10.1002/2014RS005529}
\BIBentrySTDinterwordspacing

\bibitem{Bharadwaj2012}
N.~Bharadwaj and V.~Chandrasekar, ``Wideband waveform design principles for
  solid-state weather radars,'' \emph{Journal of Atmospheric and Oceanic
  Technology}, vol.~29, no.~1, pp. 14--31, 2012.

\bibitem{Kumar8517944}
M.~{Kumar}, S.~{Joshill}, M.~{Vega}, V.~{Chandrasekar}, and J.~W. {Zebley},
  ``Nasa d3r: 2.0, enhanced radar with new data and control features,'' in
  \emph{IGARSS 2018 - 2018 IEEE International Geoscience and Remote Sensing
  Symposium}, July 2018, pp. 7978--7981.

\bibitem{8128188}
M.~{Kumar}, S.~S. {Joshil}, V.~{Chandrasekar}, R.~M. {Beauchamp}, M.~{Vega},
  and J.~W. {Zebley}, ``Performance trade-offs and upgrade of nasa d3r weather
  radar,'' in \emph{2017 IEEE International Geoscience and Remote Sensing
  Symposium (IGARSS)}, July 2017, pp. 5260--5263.

\bibitem{kumar2019receive}
M.~Kumar, Dileep, K.~Sreenivasulu, D.~Seshagiri, D.~Srinivas, and
  S.~Narasimhan, ``Receive signal path design for active phased array radars,''
  2019.

\bibitem{8899120}
V.~{Chandrasekar}, S.~S. {Joshil}, M.~{Kumar}, M.~A. {Vega}, D.~{Wolff}, and
  W.~{Petersen}, ``Snowfall observations during the winter olympics of 2018
  campaign using the d3r radar,'' in \emph{IGARSS 2019 - 2019 IEEE
  International Geoscience and Remote Sensing Symposium}, July 2019, pp.
  4561--4564.

\bibitem{Icepop1}
V.~{Chandrasekar}, M.~A. {Vega}, S.~{Joshil}, M.~{Kumar}, D.~{Wolff}, and
  W.~{Petersen}, ``Deployment and performance of the nasa d3r during the
  ice-pop 2018 field campaign in south korea,'' in \emph{IGARSS 2018 - 2018
  IEEE International Geoscience and Remote Sensing Symposium}, July 2018, pp.
  8349--8351.

\end{thebibliography}

\end{document}